\documentclass[twocolumn,10pt,conference]{IEEEtran}
\usepackage[nolist,nohyperlinks]{acronym} 
\usepackage{algorithm,algpseudocode}
\usepackage{amsmath,amssymb,amsfonts}
\usepackage{array,booktabs,multirow,longtable}
\usepackage{graphicx,caption,subcaption,xspace,xcolor,cite}
\usepackage{titlesec}

\pdfoutput=1

\IEEEoverridecommandlockouts
\def\BibTeX{{\rm B\kern-.05em{\sc i\kern-.025em b}\kern-.08em  
    T\kern-.1667em\lower.7ex\hbox{E}\kern-.125emX}}

\usepackage{tikz,tikz-3dplot,tkz-tab}
\usetikzlibrary{angles, calc, positioning, quotes, intersections,fit,decorations.pathreplacing,decorations.markings}
\tdplotsetmaincoords{70}{120}

\tikzset{
    cross/.style={fill=white,path picture={\draw[black]
        (path picture bounding box.south east) -- (path picture bounding box.north west)
        (path picture bounding box.south west) -- (path picture bounding box.north east);}},
    dressed/.style={fill=white,postaction={pattern=north east lines}},
    momentum/.style={->,semithick,yshift=5pt,shorten >=5pt,shorten <=5pt},
    loop/.style 2 args={thick,decoration={markings,mark=at position {#1} with {\arrow{<},\node[anchor=\pgfdecoratedangle-90,font=\footnotesize] {};}},postaction={decorate}},
    label/.style={thin,gray,shorten <=-1.5ex}
}

\tikzset{
    cross/.pic = {
    \draw[rotate = 45] (-#1,0) -- (#1,0);
    \draw[rotate = 45] (0,-#1) -- (0, #1);
    }
}

\begin{document}
    \bstctlcite{IEEE:BSTcontrol}

    

    \begin{acronym}
        \acro{ALS}{alternating least squares}
        \acro{AoA}{azimuth of arrival}
        \acro{AoD}{azimuth of departure}
        \acro{AWGN}{additive white Gaussian noise}
        \acro{BALS}{bilinear alternating least squares}
        \acro{BD-RIS}{beyond-diagonal RIS}
        \acro{CRLB}{Cramér-Rao lower bound}
        \acro{CSI}{channel state information}
        \acro{DFRC}{dual-function radar-communication}
        \acro{DI}{Doppler ignorant}
        \acro{EoA}{elevation of arrival}
        \acro{EoD}{elevation of departure}
        \acro{ESPRIT}{estimation of the signal parameters via rotational invariance techniques}
        \acro{EVD}{eigenvalue decomposition}
        \acro{FIM}{Fisher information matrix}
        \acro{HOSVD}{higher-order singular value decomposition}
        \acro{ISAC}{integrated sensing and communications}
        \acro{KF}{Kronecker factorization}
        \acro{KRSA}{Khatri-Rao sum approximation}
        \acro{KSA}{Kronecker sum approximation}
        \acro{LOS}{line-of-sight}
        \acro{LS}{least squares}
        \acro{MIMO}{multiple-input multiple-output}
        \acro{ML}{maximum likelihood}
        \acro{NLOS}{non-line-of-sight}
        \acro{NMSE}{normalized mean squared error}
        \acro{NTFE}{nested Tucker factorization estimation}
        \acro{OFDM}{orthogonal frequency division multiplexing}
        \acro{OMP}{orthogonal matching pursuit}
        \acro{PDF}{probability density function}
        \acro{RCS}{radar cross section}
        \acro{RIS}{reconfigurable intelligent surface}
        \acro{RMSE}{root mean squared error}
        \acro{SNR}{signal-to-noise ratio}
        \acro{SR}{sensing receiver}
        \acro{ST}{sensing transmitter}
        \acro{SVD}{singular value decomposition}
        \acro{TenDAE}{tensor Doppler-delay and angle estimation}
    \end{acronym}
    
    \title{Decoupled Delay-Doppler and Angle Estimation in BD-RIS Sensing via Nested Tucker Decomposition}
    
    \author{\IEEEauthorblockN{Kenneth Benício, André L. F. de Almeida, Fazal-E-Asim, Bruno Sokal \\ Behrooz Makki, Gabor Fodor, and A. Lee Swindlehurst}}

    \maketitle
    
     \begin{abstract}
       We study single-target localization in a group-connected beyond-diagonal reconfigurable intelligent surface (BD-RIS)-assisted monostatic network with $K$ element groups. We propose a Nested Tensor Factorization and Estimation (NTFE) algorithm that models the received signal as a $ 3$-rd-order nested Tucker tensor, thereby decoupling the delay-Doppler and angle domains. The resulting two-stage procedure estimates the target-bearing tensor factors and then extracts the other physical parameters using subspace and closed-form steps. We also analyze identifiability and uniqueness conditions. Simulations show that NTFE exploits the group-connected BD-RIS structure and outperforms state-of-the-art sensing benchmarks.
    \end{abstract}

    \begin{IEEEkeywords}
        Beyond diagonal reconfigurable surfaces, target sensing, nested tensor decomposition, Tucker decomposition. 
    \end{IEEEkeywords}

    \section{Introduction}
       \hspace*{\parindent} \Ac{RIS} technology shapes wireless propagation to improve coverage, spectral efficiency, and \ac{SNR} \cite{Rui_Zhang_2021}, but passive \ac{RIS} operation relies on cascaded-channel estimation with suitable training \cite{gong2020toward,de2021channel,Swindle2022,benicio2023tensor_wcl}. For sensing and localization at the \ac{ST}, passive \ac{BD-RIS} architectures have emerged for \ac{ISAC} \cite{Fan_2022}, using inter-element coupling to realize generally non-diagonal phase-shift matrices and extra degrees of freedom \cite{li2023reconfigurable}. Following \cite{nerini2024beyond}, we adopt a group-connected \ac{BD-RIS} that preserves intra-group coupling while avoiding full interconnection, balancing flexibility, hardware complexity, and identifiability. Related estimation, beamforming, and \ac{ISAC} applications appear in \cite{liu2024enhancing,ginige2025efficient,de2025channel,benicio2026multi}.\\
        \indent Tensor-based methods are well suited for this setting because they exploit the multilinear structure of the received signal rather than treating the channel as an unstructured matrix \cite{de2007parafac,de2008constrained,FAVIER_2012,Favier_EUSIPICO}. Our previous work used a two-stage nested tensor approach based on ESPRIT (Estimation of Signal Parameters via Rotational Invariance Techniques) for delay, Doppler, and angle estimation in diagonal-\ac{RIS}-assisted \ac{MIMO} sensing \cite{benicio2024low,benicio2024ris}, and later applied tensor processing to \ac{BD-RIS}-assisted bistatic multi-target sensing \cite{benicio2026multi}. This paper bridges these efforts by studying single-target sensing with a group-connected \ac{BD-RIS} while explicitly exploiting the block-wise structure induced by the element groups.  \\
        \indent We propose a \ac{NTFE} algorithm for target localization in a \ac{BD-RIS}-assisted network. The received signal is modeled as a $3$rd order nested Tucker tensor that separates the delay-Doppler and angle domains. The method estimates the tensor factors, extracts the parameters using subspace and closed-form steps, and admits identifiability and uniqueness conditions. Simulations against state-of-the-art benchmarks, including \cite{ercan2025ris}, show improved estimation accuracy in terms of \ac{RMSE} and \ac{NMSE}. \\
        \textit{\textbf{Notation}}: Scalars, vectors, matrices, and tensors are denoted by $a$, $\boldsymbol{a}$, $\boldsymbol{A}$, and $\boldsymbol{\mathcal{A}}$, respectively. For matrix $\boldsymbol{A}$, $\boldsymbol{A}^{*}$, $\boldsymbol{A}^{\text{T}}$, $\boldsymbol{A}^{\text{H}}$, and $\boldsymbol{A}^{\dagger}$ denote the conjugate, transpose, Hermitian transpose, and pseudoinverse. The $j$th column of $\boldsymbol{A}$ is denoted by $\boldsymbol{a}_{j}$, and $\boldsymbol{I}_{N}$ is the $N\times N$ identity. The operator D$(\cdot)$ maps a vector to a diagonal matrix. The $n$th mode product is $\boldsymbol{\mathcal{A}} \times_{n} \boldsymbol{B} = \boldsymbol{B} [\boldsymbol{\mathcal{A}}]_{(n)}$. Operators $\otimes$, $\diamond$, $\odot$, and $\oslash$ denote Kronecker, Khatri-Rao, and Hadamard products, and element-wise division, respectively.
    \section{System Model}
        \hspace*{\parindent} We consider the monostatic setup in Fig. \ref{fig:system_model_localization}, where a multi-antenna \ac{ST} is assisted by a group-connected \ac{BD-RIS} with $N=N_zN_y$ elements split into $K$ groups, with $\sum_{k=1}^{K}N_k=\bar{N}$. The direct \ac{LOS} path is blocked, so sensing occurs through the \ac{BD-RIS}-assisted echo link. The \ac{ST} uses an $L=L_zL_y$-element uniform planar array and transmits an \ac{OFDM} pulse with $Q$ subcarriers, $M$ symbols, and pilots $\boldsymbol{x}_{q,m}\in\mathbb{C}^{L\times 1}$. The received echo is\footnote{In array partitioning, the angle information is preserved while the complex gain changes across elements only in phase \cite{9868201}.}
        \begin{align}
            \begin{split}\boldsymbol{y}_{q,m,t} &= \sum_{k=1}^{K} \alpha_k \underbrace{\boldsymbol{a}(\phi_{\text{st}}, \theta_{\text{st}}) \boldsymbol{b}^{\text{T}}(\phi_{\text{ris}_{A}},\theta_{\text{ris}_{A}}) \boldsymbol{S}^{\text{T}}_{k,t} \boldsymbol{p}(\phi_{\text{ris}_{\text{D}}}, \theta_{\text{ris}_{\text{D}}})}_{\text{Target-\ac{BD-RIS}-\ac{ST} path via group }k} \\& \times\underbrace{\boldsymbol{p}^{\text{T}}(\phi_{\text{ris}_{\text{D}}}, \theta_{\text{ris}_{\text{D}}}) \boldsymbol{S}_{k,t} \boldsymbol{b}(\phi_{\text{ris}_{A}},\theta_{\text{ris}_{A}}) \boldsymbol{a}^{\text{T}}(\phi_{\text{st}}, \theta_{\text{st}})}_{\text{ST-\ac{RIS}-Target path via group }k} \\ &\times \boldsymbol{x}_{q,m} [\boldsymbol{c}(\tau)]_{q} [\boldsymbol{d}(\nu)]_{m} + \boldsymbol{z}_{q,m,t},
            \end{split}
        \end{align}
        where $\boldsymbol{a}(\phi_{\text{st}},\theta_{\text{st}})$ is the \ac{ST} response, and $\boldsymbol{b}(\phi_{\text{ris}_{A}},\theta_{\text{ris}_{A}})$ and $\boldsymbol{p}(\phi_{\text{ris}_{\text{D}}},\theta_{\text{ris}_{\text{D}}})$ are the \ac{BD-RIS} steering vectors for the \ac{ST} and target, respectively. With half-wavelength spacing and spatial frequencies $\mu_{\text{st}}=\pi\text{sin}(\phi_{\text{st}})\text{sin}(\theta_{\text{st}})$ and $\psi_{\text{st}}=\pi\text{cos}(\phi_{\text{st}})$, the \ac{ST} response is
    	\begin{align}
            \boldsymbol{a}(\mu_{\text{st}},\psi_{\text{st}}) = \boldsymbol{a}_{y}(\mu_{\text{st}}) \otimes \boldsymbol{a}_{z}(\psi_{\text{st}}) \in \mathbb{C}^{L \times 1},
        \end{align}
        where 
        \begin{align}
            \notag \boldsymbol{a}_{y}(\mu_{\text{st}}) &= [1,  e^{-j\mu_{\text{st}}}, \cdots, e^{-j  (L_{y} - 1) \mu_{\text{st}}} ]^{\text{T}} \in \mathbb{C}^{L_{y} \times 1}, \\
            \notag \boldsymbol{a}_{z}(\psi_{\text{s}t}) &= [1,  e^{-j\psi_{\text{st}}}, \cdots, e^{-j  (L_{z} - 1) \psi_{\text{st}}} ]^{\text{T}} \in \mathbb{C}^{L_{z} \times 1}.
        \end{align}
    	The \ac{BD-RIS} responses are defined analogously. 
        
\begin{figure}[!t]
            \centering
            \includegraphics[width=0.875\columnwidth]{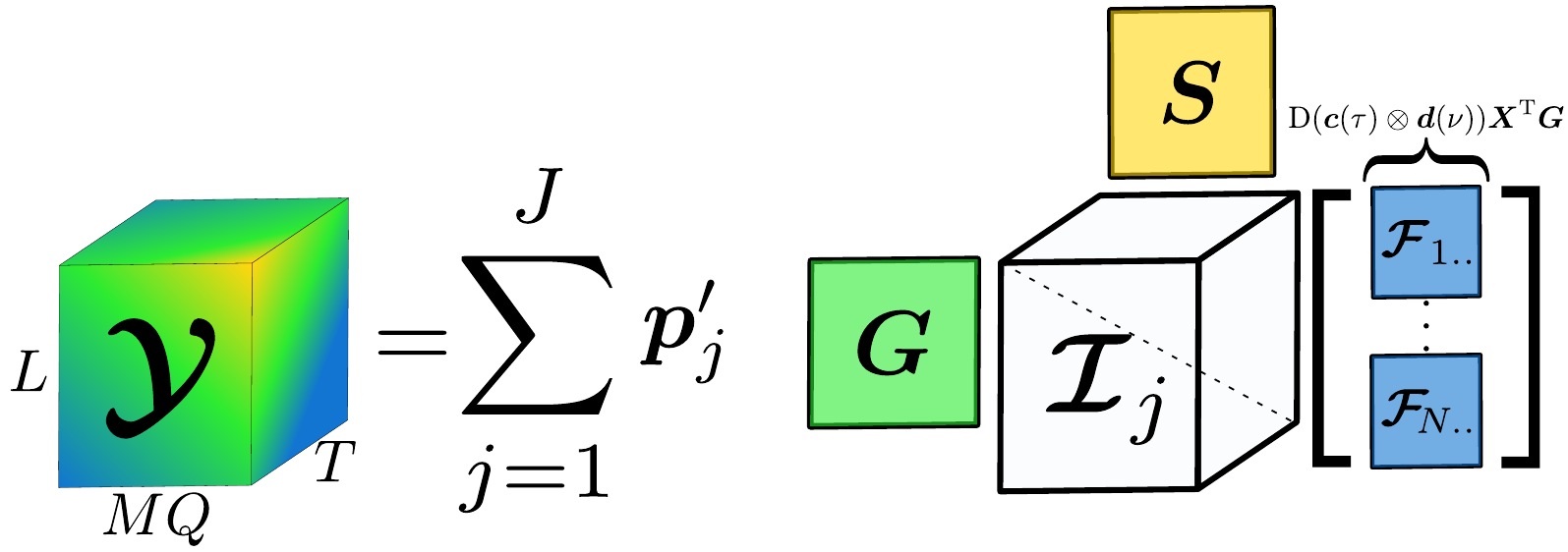}
            \caption{Noiseless nested Tucker tensor $\boldsymbol{\mathcal{Y}}$. The mode-$2$ unfolding $[\boldsymbol{\mathcal{Y}}]_{(2)}$ contains the mode-$1$ unfolding of $\hat{\boldsymbol{\mathcal{F}}}$. 
            The summation represents the mode-$4$ contraction with $\boldsymbol{p}' \in \mathbb{C}^{1 \times N^{2}}$, whose $j$th element is $p'_{j}$ for $j \in \{1, \cdots, J\}$ and $J = N^{2}$.} 
            \label{fig:nested_tucker}
            \vskip -0.3cm
        \end{figure}

        The phase-shift matrix $\boldsymbol{S}_{k,t}\in\mathbb{C}^{N_k\times N_k}$ of the $k$th group satisfies $\boldsymbol{S}^{\text{H}}_{k,t}\boldsymbol{S}_{k,t}=\boldsymbol{I}_{N_k}$, and $\boldsymbol{S}_{t}=\text{blkdiag}(\boldsymbol{S}_{1,t},\ldots,\boldsymbol{S}_{K,t})$ \cite{shen2021modeling}. The delay and Doppler steering vectors are $\boldsymbol{c}(\tau)=[1,\cdots,e^{-j2\pi(Q-1)\Delta f\tau}]^{\text{T}}$ and $\boldsymbol{d}(\nu)=[1,\cdots,e^{j2\pi(M-1)T_s\nu}]^{\text{T}}$, while $\boldsymbol{z}_{q,m,t}$ is \ac{AWGN} and $\alpha_k$ is the group-dependent complex gain \cite{benicio2026multi}. Stacking all subcarriers and symbols gives
        \begin{align}
            \begin{split}
                \boldsymbol{Y}_{t} &= \sum_{k=1}^{K} \boldsymbol{G}_{k} \boldsymbol{S}^{\text{T}}_{k,t} \boldsymbol{p}(\phi_{\text{ris}_{\text{D}}}, \theta_{\text{ris}_{\text{D}}}) \boldsymbol{p}^{\text{T}}(\phi_{\text{ris}_{\text{D}}}, \theta_{\text{ris}_{\text{D}}}) \boldsymbol{S}_{k,t} \boldsymbol{G}^{\text{T}}_{k} \\& \times \boldsymbol{X} \text{D}(\boldsymbol{c}(\tau) \otimes \boldsymbol{d}(\nu))  + \boldsymbol{Z}_{t} \in \mathbb{C}^{L \times M Q},
            \end{split}
        \end{align}
        where $\boldsymbol{G}_{k} =  \alpha_k \boldsymbol{a}(\phi_{\text{st}}, \theta_{\text{st}}) \boldsymbol{b}^{\text{T}}(\phi_{\text{ris}_{A}},\theta_{\text{ris}_{A}}) \in \mathbb{C}^{L \times N_k}$ is the geometric channel between the \ac{ST} and the $k$th \ac{BD-RIS} group, and $\boldsymbol{Z}_{t}$ is \ac{AWGN} at the $t$th time-slot. For ease of notation, we omit the noise terms in the following derivations, and consider its impact later in the numerical evaluations. Since both nodes are at fixed locations, we assume $\{\boldsymbol{G}_{k}\}_{k=1}^{K}$ is known \textit{a priori} via standard channel estimation procedures, which is a common assumption in the literature \cite{ercan2025ris,benicio2024low,benicio2024ris}. Applying the property $\text{vec}(\boldsymbol{A} \boldsymbol{B} \boldsymbol{C}) = (\boldsymbol{C}^{\text{T}} \otimes \boldsymbol{A}) \text{vec} (\boldsymbol{B})$ and defining $\boldsymbol{F}_{\tau\nu,k} = \boldsymbol{G}_{k}^{\text{T}} \boldsymbol{X} \text{D}(\boldsymbol{c}(\tau) \otimes \boldsymbol{d}(\nu)) \in \mathbb{C}^{N_k \times M Q}$ as the group-wise matrix containing the delay and Doppler information leads to 
        \begin{align}
            \notag \boldsymbol{y}_{t} &= \sum_{k=1}^{K} \{\boldsymbol{p}' \otimes (\boldsymbol{F}_{\tau\nu,k}^{\text{T}} \otimes \boldsymbol{G}_{k})\} \text{vec}((\boldsymbol{S}_{k,t} \otimes \boldsymbol{S}_{k,t})^{\text{T}}), 
        \end{align}
        where $\boldsymbol{p}'=\text{vec}^{\text{T}}(\boldsymbol{p}\boldsymbol{p}^{\text{T}})\in\mathbb{C}^{1\times N_k^2}$ and $\boldsymbol{y}_{t} = \text{vec}(\boldsymbol{Y}_{t}) \in \mathbb{C}^{L M Q \times 1}$. Assuming a specific group of size $N_{k}$ and collecting all $T$ samples of the pilot signal in a column, we obtain
        \begin{align}
             \boldsymbol{Y} &=  \sum_{k=1}^{K} \{ \boldsymbol{p} \otimes \boldsymbol{F}_{\tau\nu,k}^{\text{T}} \otimes \boldsymbol{G}_{k}\} \boldsymbol{S}^{\text{T}}_{k}, \label{eq:received_signal_all_samples}
        \end{align}
        where $\boldsymbol{S}_{k} = [\text{vec}(\boldsymbol{S}_{k,1}^{\text{T}} \otimes \boldsymbol{S}_{k,1}^{\text{T}}), \cdots, \text{vec}(\boldsymbol{S}_{k,T}^{\text{T}} \otimes \boldsymbol{S}_{k,T}^{\text{T}}) ]^{\text{T}} \in \mathbb{C}^{T \times N^{4}_{k}}$. 
        \begin{figure}
            \centering
            \begin{tikzpicture}[scale=1, every node/.style={scale=0.975}]
                \begin{scope}[
                    box1/.style={draw=black, thick, rectangle,rounded corners, minimum height=0.5cm, minimum width=0.5cm}]
                    \node (IRS 1) at (-2,0.75) {$\text{BD-RIS}$};
                    \draw[green!45!black,dashed,fill=white] (-3.5,-2.5) rectangle (-.5,.5);
                    \foreach \x/\y/\fillcolor in {-2.68/-0.28/green!18,-1.32/-0.28/lime!28,-2.68/-1.72/green!32,-1.32/-1.72/olive!25}{
                        \draw[black,thick,rounded corners,fill=\fillcolor] (\x-0.62,\y-0.62) rectangle (\x+0.62,\y+0.62);
                        \foreach \elementx in {-0.34,0,0.34}{
                            \foreach \elementy in {-0.34,0,0.34}{
                                \node[draw=green!45!black,fill=white,minimum size=0.13cm,inner sep=0pt] at (\x+\elementx,\y+\elementy) {};
                            }
                        }
                    }
                    \draw[black,fill=blue!30] (-7+1.5,-3.5) -- (-6.5+1.5,-3.5) node[above]{$\text{ST}$} -- (-6+1.5,-3.5) -- (-6.5+1.5,-2.4) -- cycle;
                    \node[circle,draw=black,fill=blue!30,minimum size=12pt] (T1) at (2.70-2,-2-1) {};
                    \node[above] (Target) at (2.70-2,-1.5-1.26) {$\text{Target}$};
                    \draw[line width=0.25mm,black] (-7+1.5,-2.4) -- (-6+1.5,-2.4) node[midway,above]{$\cdots$};
                    \draw[line width=0.25mm,black] (-7+1.5,-2.4) -- (-7+1.5,-2.35);
                    \draw[line width=0.25mm,red] (-6+1.5,-2.4) -- (-6+1.5,-2.35);
                    \draw[line width=0.25mm,black] (-7.20+1.5,-2.25) -- (-7+1.5,-2.35) -- (-6.80+1.5,-2.25) -- cycle;
                    \draw[line width=0.25mm,black] (-6.20+1.5,-2.25) -- (-6+1.5,-2.35) -- (-5.80+1.5,-2.25) -- cycle;
                    \draw[line width=0.75mm,black,->] (-6.5+1.5,-1.65) -- (-3.75,0) node[midway, above, rotate=+50]{};
                    \draw[line width=0.75mm,red,<-] (-6.5+1.5,-1.95) -- (-3.75,-0.30) node[midway, above, rotate=+50]{};
                    \draw[line width=0.75mm,black,->] (-0.25,0) -- (2-1.15,-1.65);
                    \draw[line width=0.75mm,red,<-] (-0.25,-0.30) -- (2-1.15,-1.95);
                    \draw[line width=0.75mm,dashed,black!75,->] (-6+1.5,-3) -- (-2.25,-3) node[midway, above, rotate=+50]{};
                    \draw (-1.95,-3) pic[rotate = 0] {cross=7pt};
                \end{scope}
            \end{tikzpicture}
            \caption{Group-connected BD-RIS-assisted localization.}
            \label{fig:system_model_localization}
        \end{figure}
        The preceding expressions describe the generic group-connected architecture before fixing a particular processing block. In the remainder of the derivation, we focus on an arbitrary group $k$ and perform all processing within this selected group. Thus the group index $k$ is suppressed from this point onward, so that $N$, $\boldsymbol{G}$, $\boldsymbol{S}$, and $\boldsymbol{F}_{\tau\nu}$ refer to the corresponding quantities of the selected group. This convention does not restrict the model to a fully-connected \ac{BD-RIS}, since the same processing can be applied to any group, and the overall group-connected formulation remains valid for an arbitrary number of groups. Thus, for the $k$th group, (\ref{eq:received_signal_all_samples}) corresponds to the third mode transpose of a tensor $\boldsymbol{\mathcal{Y}} \in \mathbb{C}^{L \times M Q \times T}$, which is expressed as a $3$rd order Tucker tensor decomposition multiplied by the vector containing the angle information of the $4$th mode, given by This yields the tensor $\boldsymbol{\mathcal{Y}}\in\mathbb{C}^{L\times MQ\times T}$ represented by the nested Tucker model
        \begin{align}
            \boldsymbol{\mathcal{Y}} &= \boldsymbol{\mathcal{I}} \times_{1} \boldsymbol{G} \times_{2} [\boldsymbol{\mathcal{F}}]_{(1)} \times_{3} \boldsymbol{S} \times_{4} \boldsymbol{p}', \label{eq:tensor_1}
        \end{align}
        where $\boldsymbol{\mathcal{I}} \in \mathbb{C}^{N \times N \times N^{4} \times N^{2}}$ is the known core tensor built from the identity $[\boldsymbol{\mathcal{I}}]_{(3)} = \boldsymbol{I}_{N^4}$ and  $[\boldsymbol{\mathcal{F}}]_{(1)} = \boldsymbol{F}_{\tau\nu}^{\text{T}} \in \mathbb{C}^{N \times MQ}$ is the mode-$1$ unfolding of a $3$rd order Tucker tensor. The unfoldings of $\boldsymbol{\mathcal{Y}}$ are then expressed as
        \begin{align}
            [\boldsymbol{\mathcal{Y}}]_{(1)} &= \boldsymbol{G} [\boldsymbol{\mathcal{I}}]_{(1)} (\boldsymbol{p}' \otimes \boldsymbol{S} \otimes [\boldsymbol{\mathcal{F}}]_{(1)})^{\text{T}}  \in \mathbb{C}^{L \times M Q T}, \label{eq:tensor_1_unfolding_1} \\
            [\boldsymbol{\mathcal{Y}}]_{(2)} &= [\boldsymbol{\mathcal{F}}]_{(1)} [\boldsymbol{\mathcal{I}}]_{(2)} (\boldsymbol{p}' \otimes \boldsymbol{S} \otimes \boldsymbol{G})^{\text{T}} \in \mathbb{C}^{M Q\times L T}, \label{eq:tensor_1_unfolding_2} \\
            [\boldsymbol{\mathcal{Y}}]_{(3)} &= \boldsymbol{S} [\boldsymbol{\mathcal{I}}]_{(3)} (\boldsymbol{p}' \otimes [\boldsymbol{\mathcal{F}}]_{(1)} \otimes \boldsymbol{G})^{\text{T}} \in \mathbb{C}^{T \times L M Q}, \label{eq:tensor_1_unfolding_3} \\
            [\boldsymbol{\mathcal{Y}}]_{(4)} &= \boldsymbol{p}' [\boldsymbol{\mathcal{I}}]_{(4)} (\boldsymbol{S} \otimes [\boldsymbol{\mathcal{F}}]_{(1)} \otimes \boldsymbol{G})^{\text{T}} \in \mathbb{C}^{1 \times L M Q T}. \label{eq:tensor_1_unfolding_4}
        \end{align}
        The delay-Doppler factors are further isolated through
        \begin{align}
            \hat{\boldsymbol{\mathcal{F}}} &= \boldsymbol{\mathcal{X}} \times_{1} \boldsymbol{G}^{\text{T}} \times_{2} \text{D}(\boldsymbol{d}(\nu)) \times_{3} \text{D}(\boldsymbol{c}(\tau)) \in \mathbb{C}^{N \times M \times Q}, \label{eq:tensor_2}
        \end{align}
        where $\boldsymbol{\mathcal{X}}\in\mathbb{C}^{L\times M\times Q}$ satisfies $[\boldsymbol{\mathcal{X}}]_{(1)}=\boldsymbol{X}$, with unfoldings
        \begin{align}
            [\hat{\boldsymbol{\mathcal{F}}}]_{(1)} &= \boldsymbol{G}^{\text{T}} [\boldsymbol{\mathcal{X}}]_{(1)} (\text{D}(\boldsymbol{c}(\tau)) \otimes \text{D}(\boldsymbol{d}(\nu))) \hspace{-0.05cm}\in\hspace{-0.05cm} \mathbb{C}^{N \times M Q}, \label{eq:tensor_2_unfolding_1} \\
            [\hat{\boldsymbol{\mathcal{F}}}]_{(2)} &= \text{D}(\boldsymbol{d}(\nu)) [\boldsymbol{\mathcal{X}}]_{(2)} (\text{D}(\boldsymbol{c}(\tau)) \otimes \boldsymbol{G}^{\text{T}})^{\text{T}} \hspace{-0.05cm}\in\hspace{-0.05cm} \mathbb{C}^{M \times N Q}, \label{eq:tensor_2_unfolding_2} \\ 
            [\hat{\boldsymbol{\mathcal{F}}}]_{(3)} &= \text{D}(\boldsymbol{c}(\tau)) [\boldsymbol{\mathcal{X}}]_{(3)} (\text{D}(\boldsymbol{d}(\nu)) \otimes \boldsymbol{G}^{\text{T}})^{\text{T}} \hspace{-0.05cm}\in\hspace{-0.05cm} \mathbb{C}^{Q \times N M}. \label{eq:tensor_2_unfolding_3}
        \end{align}
        \indent Equations (\ref{eq:tensor_1}) and (\ref{eq:tensor_2}) form the nested Tucker model, separating the delay-Doppler and angle parameters for structured extraction, as illustrated in Fig. \ref{fig:nested_tucker}.
        
    \section{Nested Tucker Tensor Target Sensing} \label{sec:nested_tucker_tensor_target_sensing}
        \hspace*{\parindent} Here, we propose an iterative solution for the nested Tucker tensor model expressed in (\ref{eq:tensor_1}) and (\ref{eq:tensor_2}). We fit these models with the well-known \ac{ALS} \cite{comon2009tensor,deAraujoSAM2020, benicio2023tensor}, which yields subspace estimates of the \ac{CSI} factors $[\hat{\boldsymbol{\mathcal{F}}}]_{(1)}$, and $\boldsymbol{p}'$. The parameters of interest are then extracted via subspace techniques, and the complex gain $\boldsymbol{\alpha}$ is obtained afterward through a closed-form \ac{LS} step \cite{ardah2021trice}.
        
        \subsection{Decoupling the Angle Information}
            \indent Starting from the Tucker model in (\ref{eq:tensor_1}), the \ac{ST} estimates the factor matrices that contain the target information $[\boldsymbol{\mathcal{F}}]_{(1)}$ and $\boldsymbol{p}'$, by solving the following LS fitting problem:
            \begin{equation}
                \{[\hat{\boldsymbol{\mathcal{F}}}]_{(1)}, \hat{\boldsymbol{p}'}\} = \underset{[\hat{\boldsymbol{\mathcal{F}}}]_{(1)}, \boldsymbol{p}'}{\text{arg min}} \hspace{-0.075cm} \left| \hspace{-0.05cm} \left| \begin{aligned}\boldsymbol{\mathcal{Y}} - \boldsymbol{\mathcal{I}} \times_{1} \boldsymbol{G} \times_{2} \\ [\hat{\boldsymbol{\mathcal{F}}}]_{(1)} \times_{3} \boldsymbol{S} \times_{4} \boldsymbol{p}'&\end{aligned} \right| \hspace{-0.05cm} \right|^{2}_{\text{F}}, \hspace{-0.1cm} \label{eq:tensor_fit_problem_1}
            \end{equation}
            which can be solved by employing the \ac{BALS} procedure, which consists of alternatingly solving the following optimization problems until a stopping criterion is reached\footnote{The convergence criteria for the proposed algorithms are specified in Alg.~\ref{alg:proposed_1}. Termination typically occurs upon reaching a predefined maximum number of iterations or when the objective function satisfies a convergence tolerance between consecutive iterations.}:
            \begin{align}
                [\hat{\boldsymbol{\mathcal{F}}}]_{(1)} &= \underset{[\boldsymbol{\mathcal{F}}]_{(1)}}{\text{arg min}} \left\| \begin{aligned} & \left[\boldsymbol{\mathcal{Y}}\right]_{(2)} - [\boldsymbol{\mathcal{F}}]_{(1)} [\boldsymbol{\mathcal{I}}]_{(2)} (\boldsymbol{p}' \otimes \boldsymbol{S} \otimes \boldsymbol{G})^{\text{T}} \end{aligned} \right\|^{2}_{\text{F}}, \nonumber \\ 
                &= [\boldsymbol{\mathcal{Y}}]_{(2)} [[\boldsymbol{\mathcal{I}}]_{(2)} (\boldsymbol{p}' \otimes \boldsymbol{S} \otimes \boldsymbol{G})^{\text{T}}]^{\dagger}, \label{eq:tensor_1_pseudoinverse_1} \\
                \hat{\boldsymbol{p}'} &= \underset{\boldsymbol{p}'}{\text{arg min}} \left\| \begin{aligned} & 
                \left[\boldsymbol{\mathcal{Y}}\right]_{(4)} - \boldsymbol{p}' [\boldsymbol{\mathcal{I}}]_{(4)} ( \boldsymbol{S} \otimes [\hat{\boldsymbol{\mathcal{F}}}]_{(1)} \otimes \boldsymbol{G} )^{\text{T}} \end{aligned} 
                \right\|^{2}_{\text{F}}, \nonumber \\
                &= [\boldsymbol{\mathcal{Y}}]_{(4)} [[\boldsymbol{\mathcal{I}}]_{(4)} ([\hat{\boldsymbol{\mathcal{F}}}]_{(1)} \otimes \boldsymbol{G} \otimes \boldsymbol{S})^{\text{T}}]^{\dagger}. \label{eq:tensor_1_pseudoinverse_2}
            \end{align}
            
            The solutions above exist as long as $LT \geq N$ and $L M Q T \geq N^{2}$, respectively. With the above, we have estimates of the mode-$1$ unfoldings of the Tucker tensor $\hat{\boldsymbol{\mathcal{F}}}$ that contains the target delay and Doppler information. 
            
        \subsection{Decoupled Delay and Doppler Information Estimation} 
            \indent Using an approach similar to the first stage, we estimate the Tucker model in (\ref{eq:tensor_2}), which contains the decoupled delay and Doppler information $\boldsymbol{d}(\nu)$ and $\boldsymbol{c}(\tau)$, by solving the following LS fitting problem
            \begin{equation}
                \hspace{-0.3cm} \{\hat{\boldsymbol{d}}(\nu), \hat{\boldsymbol{c}}(\tau)\} \hspace{-0.05cm}=\hspace{-0.05cm} \underset{\boldsymbol{d}(\nu), \boldsymbol{c}(\tau)}{\text{arg min}} \left|\left|\begin{aligned}\hat{\boldsymbol{\mathcal{F}}} - \boldsymbol{\mathcal{X}} \times_{1} \boldsymbol{G}^{\text{T}} \times_{2}& \\ \text{D}(\boldsymbol{d}(\nu)) \times_{3} \text{D}(\boldsymbol{c}(\tau))&\end{aligned}\right|\right|^{2}_{\text{F}}. \label{eq:tensor_fit_problem_2}
            \end{equation}
            \indent Applying the property $\text{vec}(\boldsymbol{A} \text{D}(\boldsymbol{b}) \boldsymbol{C}) = (\boldsymbol{C}^{\text{T}} \diamond \boldsymbol{A})\boldsymbol{b}$ to (\ref{eq:tensor_2_unfolding_2}) and (\ref{eq:tensor_2_unfolding_3}), which structurally enforces diagonalization and prevents noise from inducing erroneous non-zero off-diagonal elements, we have
            \begin{align}
                \hspace{-0.1cm}\text{vec}([\hat{\boldsymbol{\mathcal{F}}}]_{(2)}) &\hspace{-0.05cm}=\hspace{-0.05cm} [\hspace{-0.05cm}([\boldsymbol{\mathcal{X}}]_{(2)} \text{D}(\boldsymbol{c}(\tau)) \hspace{-0.05cm}\otimes\hspace{-0.05cm} \boldsymbol{G})^{\text{T}} \hspace{-0.15cm}\diamond\hspace{-0.05cm} \boldsymbol{I}_{M}\hspace{-0.05cm} ] \boldsymbol{d}(\tau), \label{eq:tensor_2_unfolding_1_vec} \\
                \hspace{-0.1cm}\text{vec}([\hat{\boldsymbol{\mathcal{F}}}]_{(3)}) &\hspace{-0.05cm}=\hspace{-0.05cm} [\hspace{-0.05cm}([\boldsymbol{\mathcal{X}}]_{(3)} \text{D}(\boldsymbol{d}(\nu)) \hspace{-0.05cm}\otimes\hspace{-0.05cm} \boldsymbol{G})^{\text{T}} \hspace{-0.15cm}\diamond\hspace{-0.05cm} \boldsymbol{I}_{Q} \hspace{-0.05cm}] \boldsymbol{c}(\tau). \label{eq:tensor_2_unfolding_2_vec}
            \end{align}
            Employing BALS, we alternately solve the following:
            \begin{align}
                \hat{\boldsymbol{d}}(\nu) &= \underset{\boldsymbol{d}(\nu)}{\text{arg min}} \left|\left|\begin{aligned}\text{vec}([\hat{\boldsymbol{\mathcal{F}}}]_{(2)}) - [([\boldsymbol{\mathcal{X}}]_{(2)}& \\ \text{D}(\boldsymbol{c}(\tau)) \otimes \boldsymbol{G})^{\text{T}} \diamond \boldsymbol{I}_{M}] \boldsymbol{d}(\tau)& \end{aligned}\right|\right|^{2}_{\text{F}} \label{eq:stage_2_tucker_1_ls_1} \\
                &= [([\boldsymbol{\mathcal{X}}]_{(2)} \text{D}(\boldsymbol{c}(\tau)) \otimes \boldsymbol{G})^{\text{T}} \diamond \boldsymbol{I}_{M}]^{\dagger} \text{vec}([\hat{\boldsymbol{\mathcal{F}}}]_{(2)}), \label{eq:tensor_2_pseudoinverse_1} \\          
                \hat{\boldsymbol{c}}(\tau) &= \underset{\boldsymbol{c}(\tau)}{\text{arg min}} \left|\left|\begin{aligned}\text{vec}([\hat{\boldsymbol{\mathcal{F}}}]_{(3)}) - [([\boldsymbol{\mathcal{X}}]_{(3)}& \\ \text{D}(\boldsymbol{d}(\nu)) \otimes \boldsymbol{G})^{\text{T}} \diamond \boldsymbol{I}_{Q}] \boldsymbol{c}(\tau)& \end{aligned}\right|\right|^{2}_{\text{F}} \label{eq:stage_2_tucker_2_ls_2} \\
                &= [([\boldsymbol{\mathcal{X}}]_{(3)} \text{D}(\boldsymbol{d}(\nu)) \otimes \boldsymbol{G})^{\text{T}} \diamond \boldsymbol{I}_{Q}]^{\dagger} \text{vec}([\hat{\boldsymbol{\mathcal{F}}}]_{(3)}). \label{eq:tensor_2_pseudoinverse_2}
            \end{align}
            The solutions above exist as long as $N Q \geq 1$ and $N M \geq 1$, respectively. After obtaining the estimates $ \hat{\boldsymbol{d}}(\nu)$ and $ \hat{\boldsymbol{c}}(\tau)$, we apply ESPRIT to acquire estimates of the delay and Doppler, $\tau$ and $\nu$, respectively.
            
        \subsection{Angle Estimation}       
            Having estimated the delay $\hat{\tau}$ and Doppler $\hat{\nu}$ parameters, we parametrically reconstruct their respective steering vectors to refine the estimates, yielding
            \begin{align}
                \boldsymbol{c}(\hat{\tau}) &= [1, \dots, e^{-j 2 \pi (Q - 1) \Delta f \hat{\tau}} ]^{\text{T}}, \\
                \boldsymbol{d}(\hat{\nu})  &= [1, \dots, e^{j 2 \pi (M - 1) T_{s} \hat{\nu}}]^{\text{T}},
            \end{align}
            and reconstruct the mode-1 unfolding of the tensor $\boldsymbol{\mathcal{F}}$:
            \begin{align}
                [\hat{\boldsymbol{\mathcal{F}}'}]_{(1)} = \boldsymbol{G}^{\text{T}} [\boldsymbol{\mathcal{X}}]_{(1)} (\text{D}(\boldsymbol{c}(\hat{\tau})) \otimes \text{D}(\boldsymbol{d}(\hat{\nu}))).
            \end{align}
            Finally, the target's angle information matrix $\hat{\boldsymbol{P}}$ is extracted by substituting these refined tensor factors into the mode-4 unfolding of $\boldsymbol{\mathcal{Y}}$, which yields
            \begin{align}
                \hat{\boldsymbol{P}} = \text{unvec}_{N \times N}\left([\boldsymbol{\mathcal{Y}}]_{(4)} [[\boldsymbol{\mathcal{I}}]_{(4)} ([\hat{\boldsymbol{\mathcal{F}}'}]_{(1)} \otimes \boldsymbol{G} \otimes \boldsymbol{S})^{\text{T}}]^{\dagger}\right). \label{eq:angular_estimation}
            \end{align}
            From this matrix, we apply the 2D ESPRIT algorithm to estimate the azimuth and elevation angles, $\phi_{\text{ris}_{\text{D}}}$ and $\theta_{\text{ris}_{\text{D}}}$, respectively. Once these angles are obtained, we parametrically reconstruct the angle matrix containing the target's information to yield the refined estimate 
            \begin{align}
                \hat{\boldsymbol{P}'} = \boldsymbol{p}(\hat{\phi}_{\text{ris}_{\text{D}}}, \hat{\theta}_{\text{ris}_{\text{D}}}) \boldsymbol{p}^{\text{T}}(\hat{\phi}_{\text{ris}_{\text{D}}}, \hat{\theta}_{\text{ris}_{\text{D}}}). 
            \end{align}
            
        \subsection{Closed-form Complex Channel Coefficient Estimation}
            In the final step, we estimate the complex channel gain coefficient $\alpha$. Like other tensor algorithms, the parametric reconstructions above eliminate arbitrary scaling factors, and the only scalar ambiguity remaining in the reconstructed tensor signal is the channel coefficient itself. We first reconstruct the mode-$1$ unfolding of the tensor using the refined parameters:
            \begin{align}
                [\boldsymbol{\mathcal{Y}'}]_{(1)} = \boldsymbol{G} [\boldsymbol{\mathcal{I}}]_{(1)} (\text{vec}(\hat{\boldsymbol{P}'}) \otimes \boldsymbol{S} \otimes [\hat{\boldsymbol{\mathcal{F}'}}]_{(1)})^{\text{T}}.
            \end{align}
            The coefficient $\alpha$ is then extracted via a low-complexity element-wise division step, given by
            \begin{align}
                \hat{\alpha} = \mathbb{E} \left\{[\boldsymbol{\mathcal{Y}}]_{(1)} \oslash [\boldsymbol{\mathcal{Y}'}]_{(1)}\right\}. \label{eq:pathloss_estimation}
            \end{align}
            We take the sample mean (denoted here by $\mathbb{E}\{\cdot\}$) over all $LMQT$ available elements in the unfolded tensor, effectively exploiting the full multidimensional dataset to mitigate noise.
            A summary of the proposed solution is given in Alg. \ref{alg:proposed_1}.
            \begin{figure}[!t]
                \centering
                \includegraphics[width=0.975\columnwidth]{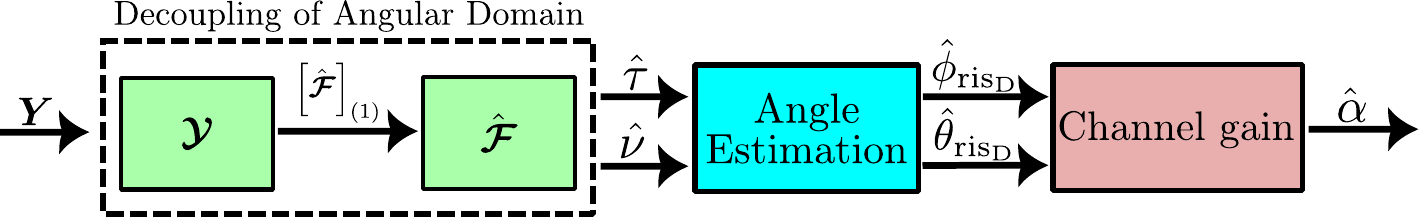}
                \caption{Block diagram of the proposed \ac{NTFE} algorithm. The nested Tucker model in (\ref{eq:tensor_1}) is solved to estimate delay and Doppler via ESPRIT, the tensor factors are refined to extract the angle parameters using (\ref{eq:angular_estimation}), and the complex channel gain is obtained in closed form via (\ref{eq:pathloss_estimation}).}
                \label{fig:solution_diagram}
            \end{figure}
            
            \begin{algorithm}[!t]
                \caption{\Acf{NTFE}} 
                \label{alg:proposed_1}
                \begin{algorithmic}[1]
                    \Require{Tensor $\boldsymbol{\mathcal{Y}}$, $\boldsymbol{G}$, $\boldsymbol{S}$, $\boldsymbol{\mathcal{X}}$, maximum number of iterations $i_{\text{max}}$, and convergence threshold $\delta$.}   
                    \State{Randomly initialize $[\hat{\boldsymbol{\mathcal{F}}}]_{(1)}$ and $\hat{\boldsymbol{p}'}$ at iteration $i = 0$.}
                    \While{$||e(i) - e(i-1)|| \geq \delta$ and $i < i_{\text{max}}$}
                    \State{Find the \ac{LS} estimate of $\boldsymbol{F}_{\tau\nu}$: $[\hat{\boldsymbol{\mathcal{F}}}]_{(1)} = [\boldsymbol{\mathcal{Y}}]_{(2)} [[\boldsymbol{\mathcal{I}}]_{(2)} (\hat{\boldsymbol{p}}' \otimes \boldsymbol{S} \otimes \boldsymbol{G})^{\text{T}}]^{\dagger}$.}
                    \State{Find the \ac{LS} estimate of $\boldsymbol{p}'$: $\hat{\boldsymbol{p}'} = [\boldsymbol{\mathcal{Y}}]_{(4)} [[\boldsymbol{\mathcal{I}}]_{(4)} ([\hat{\boldsymbol{\mathcal{F}}}]_{(1)} \otimes \boldsymbol{G} \otimes \boldsymbol{S})^{\text{T}}]^{\dagger}$.}
                    \State{Update $e(i) = ||\hat{\boldsymbol{\mathcal{Y}}} - \hat{\boldsymbol{\mathcal{Y}}}(i)||^{2}_{\text{F}}$.}
                    \EndWhile
                    \State{\textbf{return} $[\hat{\boldsymbol{\mathcal{F}}}]_{(1)}$ and $\hat{\boldsymbol{p}'}$.}
                    \State{Randomly initialize $\hat{\boldsymbol{d}}(\nu)$ and $\hat{\boldsymbol{c}}(\tau)$ at iteration $i = 0$.}
                    \While{$||e(i) - e(i-1)|| \geq \delta$ and $i < i_{\text{max}}$}
                    \State{Find the \ac{LS} estimate of $\boldsymbol{d}(\nu)$: $\hat{\boldsymbol{d}}(\nu) = [([\boldsymbol{\mathcal{X}}]_{(2)} \text{D}(\boldsymbol{c}(\tau)) \otimes \boldsymbol{G})^{\text{T}} \diamond \boldsymbol{I}_{M}]^{\dagger} \text{vec}([\hat{\boldsymbol{\mathcal{F}}}]_{(2)})$.}
                    \State{Find the \ac{LS} estimate of $\boldsymbol{c}(\tau)$: $\hat{\boldsymbol{c}}(\tau) = [([\boldsymbol{\mathcal{X}}]_{(3)} \text{D}(\boldsymbol{d}(\nu)) \otimes \boldsymbol{G})^{\text{T}} \diamond \boldsymbol{I}_{Q}]^{\dagger} \text{vec}([\hat{\boldsymbol{\mathcal{F}}}]_{(3)})$.}
                    \State{Update $e(i) = ||\hat{\boldsymbol{\mathcal{F}}} - \hat{\boldsymbol{\mathcal{F}}}(i)||^{2}_{\text{F}}$.}
                    \EndWhile
                    \State{\textbf{return} $\hat{\boldsymbol{d}}(\nu)$ and $\hat{\boldsymbol{c}}(\tau)$.}
                    \State{Extract delay $\hat{\tau}$ and Doppler $\hat{\nu}$ via 1D ESPRIT applied to $\hat{\boldsymbol{c}}(\tau)$ and $\hat{\boldsymbol{d}}(\nu)$ and parametrically reconstruct the vectors $\boldsymbol{c}(\hat{\tau}) = [1, \dots, e^{-j 2 \pi (Q - 1) \Delta f \hat{\tau}} ]^{\text{T}}$ and $\boldsymbol{d}(\hat{\nu})  = [1, \dots, e^{j 2 \pi (M - 1) T_{s} \hat{\nu}}]^{\text{T}}$.}
                    \State{Reconstruct tensor factor $[\hat{\boldsymbol{\mathcal{F}}'}]_{(1)} = \boldsymbol{G}^{\text{T}} [\boldsymbol{\mathcal{X}}]_{(1)} (\text{D}(\boldsymbol{c}(\hat{\tau})) \otimes \text{D}(\boldsymbol{d}(\hat{\nu})))$ and estimate the angle matrix: $\hat{\boldsymbol{P}} = \text{unvec}_{N \times N}([\boldsymbol{\mathcal{Y}}]_{(4)} [[\boldsymbol{\mathcal{I}}]_{(4)} ([\hat{\boldsymbol{\mathcal{F}}'}]_{(1)} \otimes \boldsymbol{G} \otimes \boldsymbol{S})^{\text{T}}]^{\dagger})$.}
                    \State{Extract angles $\phi_{\text{ris}_{\text{D}}}$ and $\theta_{\text{ris}_{\text{D}}}$ via 2D ESPRIT applied to $\hat{\boldsymbol{P}}$ and parametrically reconstruct the angle matrix $\hat{\boldsymbol{P}'} = \boldsymbol{p}(\hat{\phi}_{\text{ris}_{\text{D}}},\hat{\theta}_{\text{ris}_{\text{D}}}) \boldsymbol{p}^{\text{T}}(\hat{\phi}_{\text{ris}_{\text{D}}},\hat{\theta}_{\text{ris}_{\text{D}}})$.}
                    \State{Reconstruct received tensor signal $[\boldsymbol{\mathcal{Y}'}]_{(1)} = \boldsymbol{G} [\boldsymbol{\mathcal{I}}]_{(1)} (\text{vec}(\hat{\boldsymbol{P}'}) \otimes \boldsymbol{S} \otimes [\hat{\boldsymbol{\mathcal{F}'}}]_{(1)})^{\text{T}}$ and estimate the complex channel coefficient: $\hat{\alpha} = \mathbb{E} \{[\boldsymbol{\mathcal{Y}}]_{(1)} \oslash [\boldsymbol{\mathcal{Y}'}]_{(1)}\}$.}
                    \State{\textbf{return} $\hat{\tau}$, $\hat{\nu}$, $\hat{\phi}_{\text{ris}_{\text{D}}}$, $\hat{\theta}_{\text{ris}_{\text{D}}}$, and $\hat{\alpha}$.}
                \end{algorithmic}
            \end{algorithm}
 
    \section{Identifiability and Uniqueness} 
        \hspace*{\parindent} Unlike a standard Tucker decomposition with an unknown dense core, the proposed model uses a deterministic known core $\boldsymbol{\mathcal{I}}$ with the factor matrices corresponding to the first and third modes, which define the part of the system geometry assumed to be known at the receiver. Hence, for (\ref{eq:tensor_1}), arbitrary rotations of the estimates do not occur and the remaining ambiguities reduce to trivial scalings. Thus, the \ac{BALS} estimates satisfy $[\hat{\boldsymbol{\mathcal{F}}}]_{(1)} = [\boldsymbol{\mathcal{F}}]_{(1)} \boldsymbol{\Lambda}_{1}$ and $\hat{\boldsymbol{p}}' = \boldsymbol{p}' \boldsymbol{\Lambda}_{2}$, with $\boldsymbol{\Lambda}_{1} \in \mathbb{C}^{M Q \times M Q}$ and $\boldsymbol{\Lambda}_{2} \in \mathbb{C}^{N^{2} \times N^{2}}$ such that
        \begin{align}
        \boldsymbol{\Lambda}_{2} \otimes \boldsymbol{\Lambda}_{1} \otimes \boldsymbol{I}_{N} &= \boldsymbol{I}_{N^{4}}.
        \end{align}
        These scalings cancel each other in the reconstructed tensor and preserve the column space of $[\hat{\boldsymbol{\mathcal{F}}}]_{(1)}$, enabling delay-Doppler extraction. However, a non-uniform scaling may destroy the shift-invariance of $\hat{\boldsymbol{p}}'$, so (\ref{eq:angular_estimation}) is used to refine the angle subspace estimate. For (\ref{eq:tensor_2}), the ambiguities reduce to $\hat{\boldsymbol{d}}(\nu) = \lambda_{3} \boldsymbol{d}(\nu)$ and $\hat{\boldsymbol{c}}(\tau) = \lambda_{4} \boldsymbol{c}(\tau)$, with $\lambda_{3} \lambda_{4} = 1$. \\
        \indent Identifiability follows from the invertibility of the \ac{LS} systems in Alg. \ref{alg:proposed_1}. For (\ref{eq:tensor_1}), define $\boldsymbol{P}_{1} =  [\boldsymbol{\mathcal{I}}]_{(2)} (\boldsymbol{p}' \otimes \boldsymbol{S} \otimes \boldsymbol{G})^{\text{T}} \in \mathbb{C}^{N \times L T}$ and $\boldsymbol{P}_{2} = [\boldsymbol{\mathcal{I}}]_{(4)} ([\hat{\boldsymbol{\mathcal{F}}}]_{(1)} \otimes \boldsymbol{G} \otimes \boldsymbol{S})^{\text{T}} \in \mathbb{C}^{N^{2} \times L M Q T}$. Their right-invertibility requires $LT \geq N$ and $L M Q T \geq N^{2}$. For (\ref{eq:tensor_2}), $\boldsymbol{P}_{3} =  ([\boldsymbol{\mathcal{X}}]_{(2)} \text{D}(\boldsymbol{c}(\tau)) \otimes \boldsymbol{G})^{\text{T}} \diamond \boldsymbol{I}_{M}$ and $\boldsymbol{P}_{4} = ([\boldsymbol{\mathcal{X}}]_{(3)} \text{D}(\boldsymbol{d}(\nu)) \otimes \boldsymbol{G})^{\text{T}} \diamond \boldsymbol{I}_{Q}$ require $N Q \geq 1$, $N M \geq 1$, and $M Q \geq K$ (with $Q,M\geq 1$). Since the known core prevents arbitrary non-singular rotations \cite{benicio2023tensor}, these conditions ensure uniqueness up to trivial scaling ambiguities.
        
        \section{Simulation Results}
\begin{figure}
                \centering
                \includegraphics[width=0.60\columnwidth]{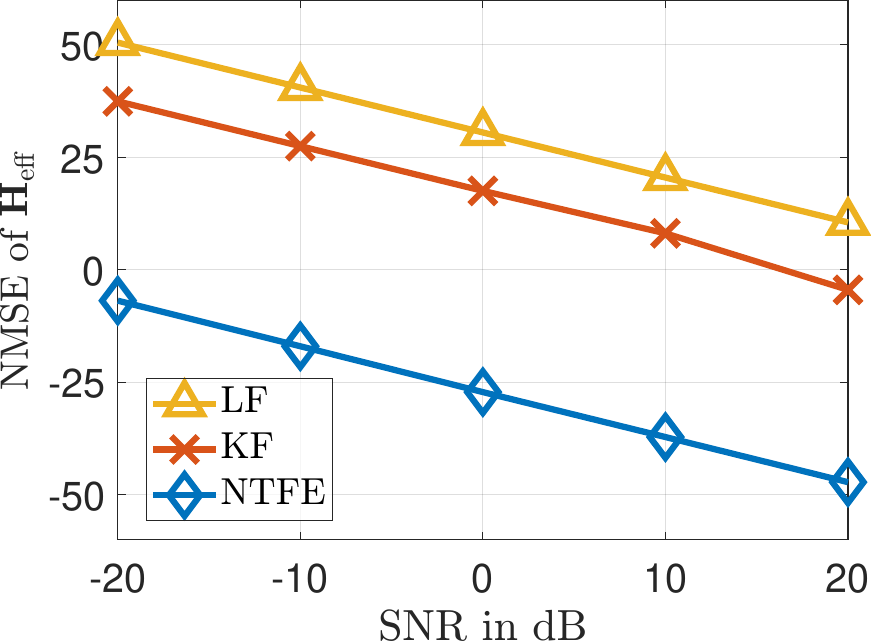}  
                \caption{\ac{NMSE} of $\boldsymbol{H}_{\text{eff}}$ estimation for \ac{NTFE}, \ac{LS}, and \ac{KF} \cite{van2000ubiquitous}.} 
                \label{fig:nmse_case_1}
            \end{figure}

\begin{figure}
            \setlength{\abovecaptionskip}{5pt}
            \setlength{\belowcaptionskip}{-1pt}
            \begin{minipage}{0.45\columnwidth}
                \includegraphics[width=\textwidth]{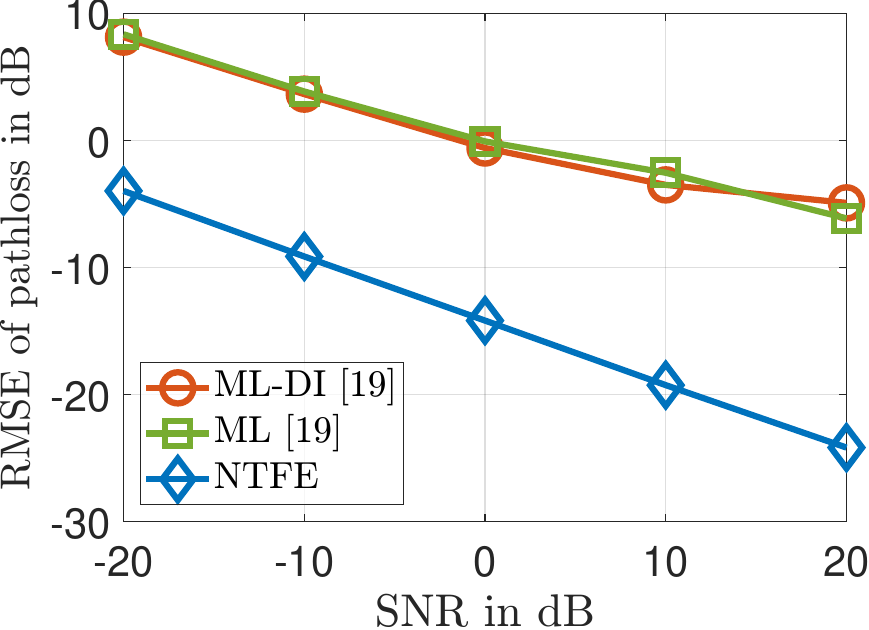}
                \caption*{(a) Complex gain.}  
            \end{minipage}
            \hfill
            \begin{minipage}{0.45\columnwidth}  
                \includegraphics[width=\textwidth]{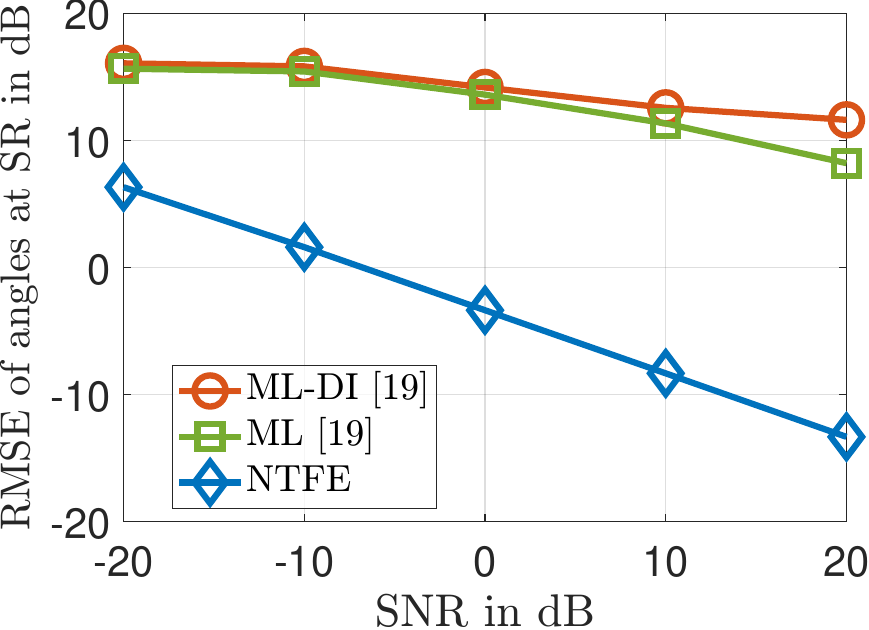}
                \caption*{(b) Angle.}   
            \end{minipage}
            \begin{minipage}{0.45\columnwidth}   
                \includegraphics[width=\textwidth]{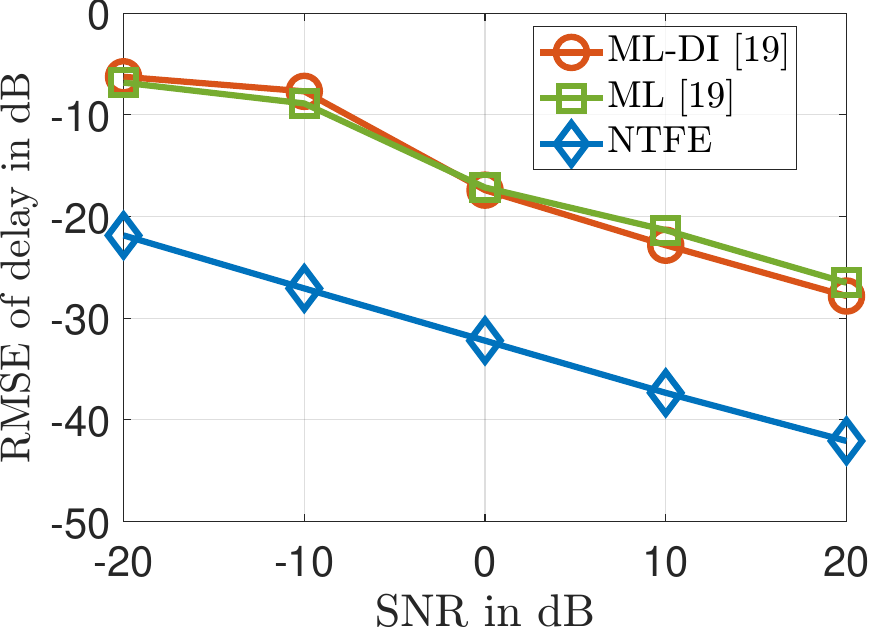}
                \caption*{(c) Delay.}
            \end{minipage}
            \hfill
            \begin{minipage}{0.45\columnwidth}   
                \includegraphics[width=\textwidth]{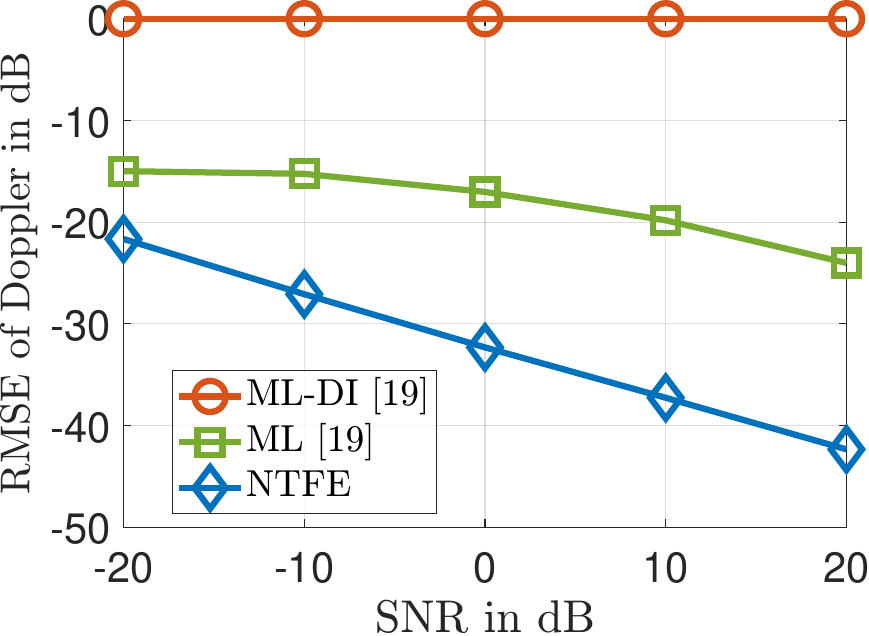}
                \caption*{(d) Doppler.}  
            \end{minipage}
            \caption{Parameter \ac{RMSE} of the proposed \ac{NTFE} algorithm.} 
            \label{fig:rmse_case_1}
            \end{figure}

            \hspace*{\parindent} We evaluate the proposed \ac{NTFE} algorithm in terms of \ac{NMSE} and \ac{RMSE} for a monostatic \ac{OFDM} sensing scenario assisted by a group-connected \ac{BD-RIS}. The reported processing considers one group with $N = N_y \times N_z$ reflecting elements, uniform planar arrays at the \ac{ST}, far-field propagation, random unitary intra-group phase-shift matrices fixed over blocks of $M Q$ \ac{OFDM} symbols, Hadamard pilots, and uniformly distributed target angles in $(0^{\circ},90^{\circ})$. Channel and parameter errors are measured by $\text{NMSE}(\boldsymbol{X}) = \mathbb{E}\{||\boldsymbol{X} - \hat{\boldsymbol{X}}||^{2}_{\text{F}}/||\boldsymbol{X}||^{2}_{\text{F}}\}$ and $\text{RMSE}(\boldsymbol{x}) = \sqrt{\mathbb{E}\{||\boldsymbol{x} - \hat{\boldsymbol{x}}||^{2}_{2}\}}$, respectively. The \ac{SNR} is defined as $\text{SNR} = ||\mathcal{\boldsymbol{Y}}||^{2}_{\text{F}}\sigma^{2}_{\mathcal{\boldsymbol{Z}}}/||\mathcal{\boldsymbol{Z}}||^{2}_{\text{F}}$, and the delay and Doppler estimates are normalized by $T_s$. Unless otherwise stated, the parameters in Table \ref{tab:parameters} are used, and all curves are averaged over $5000$ Monte Carlo realizations. We compare \ac{NTFE} with three benchmarks: the sequential $4$D \ac{ML} method of \cite{ercan2025ris}, a direct \ac{LS} estimate of (\ref{eq:received_signal_all_samples}), and the \ac{KF}-based method of\cite{van2000ubiquitous} that factorizes the Kronecker product in (\ref{eq:received_signal_all_samples}). The latter serves as a performance upper-bound when the internal spatial and temporal structures are not exploited. 
                        \begin{table}[!t] 
                \centering
                \caption{Simulation parameters.}
                \label{tab:parameters}
                \resizebox{0.675\columnwidth}{!}{
                \begin{tabular}{|c|c|}
                    \hline
                    \acs{RIS} elements & $N_{y} N_{z} = 2 \times 2 = 4$ \\ \hline
                    Antennas at the \ac{ST} & $L_{\text{ST}_{y}} L_{\text{ST}_{z}} = 2 \times 2 = 4$ \\ \hline
                    Arrays spacing & $\lambda/2$ \\ \hline
                    Angles & $\mathcal{U}(0^{\circ}, 90^{\circ})$ \\ \hline
                    \acs{ST} - \acs{RIS} & $\mathcal{U}(10\text{m},250\text{m})$ \\ \hline
                    \acs{RIS} - Cluster & $\mathcal{U}(10\text{m},250\text{m})$\\ \hline
                    Velocity & $\mathcal{U}(-25\text{m/s},+25\text{m/s})$ \\ \hline
                    \acs{RCS} & $2 \text{m}^{2}$ \\ \hline
                    Symbol duration & $1/\Delta f$ \\ \hline
                    Carrier frequency & $28$ GHz \\ \hline
                    Subcarrier spacing $\Delta f$ & $120$ KHz \\ \hline
                    Wavelength & $1.07 \times 10^{-2}$ m \\ \hline
                    $i_{\text{max}}$ & $500$ \\ \hline
                    $\delta$ & $10^{-6}$ \\ \hline
                    Symbols $M$ & $4$ \\ \hline
                    Subcarriers $Q$ & $4$ \\ \hline
                    Time-slots $T$ & $256$ \\ \hline
                \end{tabular}}
            \end{table}
            Fig.~\ref{fig:nmse_case_1} shows the \ac{NMSE} for estimating $\boldsymbol{H}_{\text{eff}} = \{ \boldsymbol{p} \otimes \boldsymbol{F}_{\tau\nu}^{\text{T}} \otimes \boldsymbol{G}\}^{\text{T}}$, reconstructed from the parameters estimated by \ac{NTFE}. The direct \ac{LS} estimate, $\hat{\boldsymbol{H}}_{\text{eff}} = \boldsymbol{S}^{\dagger} \boldsymbol{Y}$, requires $T \geq N^4$ and performs about $10$~dB worse than \ac{KF}, which benefits from the Kronecker structure. In turn, \ac{KF} remains roughly $25$~dB worse than \ac{NTFE}, since it does not exploit the full geometric structure of the sensing model. Fig. \ref{fig:rmse_case_1} reports the parameter \ac{RMSE}. By decoupling the delay-Doppler and angle domains, \ac{NTFE} accurately estimates the temporal parameters and separates the azimuth and elevation components induced by the \ac{BD-RIS}. The complex-gain accuracy follows the quality of the preceding spatial and temporal estimates. \\
            \indent \ac{NTFE} achieves a performance improvement of 10~dB compared with the standard $4$D \ac{ML} benchmark for all estimated parameters. This gain is mainly due to the reduced dependence on delay-Doppler initialization and the mitigation of error propagation from sequential \ac{ML} searches. We also show the performance of the \ac{DI} \ac{ML} variant from \cite{ercan2025ris}; although it does not estimate Doppler, it performs similarly to the standard \ac{ML} method on the remaining parameters in the considered scenario.
    \section{Conclusion}
        \hspace*{\parindent}We proposed the \ac{NTFE} algorithm for single-target localization in a group-connected \ac{BD-RIS}-assisted monostatic sensing scenario. Using a $3$rd order nested Tucker model, the method decouples the temporal and spatial parameters, supports identifiable estimation up to trivial ambiguities, and outperforms state-of-the-art benchmarks in terms of effective-channel \ac{NMSE} and parameter \ac{RMSE}.

    \renewcommand{\baselinestretch}{0.93}
      
    \bibliographystyle{IEEEtran}
    \bibliography{IEEEexample}
    
\end{document}